\documentclass[10pt,twocolumn,letterpaper,twocolumn]{article}
\usepackage{ol2}
\usepackage[draft]{hyperref}
\usepackage{amsmath}
\usepackage{calc}
\usepackage{amsfonts}
\usepackage{amssymb}
\usepackage{graphicx}
\usepackage{bm}
\usepackage{color}
\usepackage{epsfig}
\usepackage{ifthen}

\def\dz{\partial_z}
\def\M{\mathbf{M}}

\begin{document}

\twocolumn[
\title{Broadband sum frequency generation via chirped quasi-phase-matching}

\author{Andon A. Rangelov$^*$ and Nikolay V. Vitanov}

\address{
Department of Physics, Sofia University, 5 James Bourchier blvd., 1164 Sofia, Bulgaria\\
$^*$Corresponding author: rangelov@phys.uni-sofia.bg}

\begin{abstract}
An efficient broadband sum frequency generation (SFG) technique using the two collinear optical parametric processes $\omega_3=\omega_1+\omega_2$ and $\omega_4=\omega_1+\omega_3$ is proposed.
The technique uses chirped quasi-phase-matched gratings,
 which, in the undepleted pump approximation, make SFG analogous to adiabatic population transfer in three-state systems with crossing energies in quantum physics.
If the local modulation period %for aperiodically poled quasi-phase-matching
 first makes the phase match occur for $\omega_3$ and then for $\omega_4$ SFG processes
 then the energy is converted adiabatically to the $\omega_4$ field.
Efficient SFG of the $\omega_4$ field is also possible by the opposite direction of the local modulation sweep;
 then transient SFG of the $\omega_3$ field is strongly reduced.
Most of these features remain valid in the nonlinear regime of depleted pump.
\end{abstract}


 ] %% activate for two-column option

\textbf{Introduction.} Recent advances in quasi-phase-matching
(QPM) techniques \cite{Saltiel,Arie} have drawn analogies between
optical parametric processes and two- and three-state quantum
systems \cite{Longhi,Suchowski2008,Suchowski2009}. By using an
analogy to stimulated Raman adiabatic passage (STIRAP) in atomic
physics
\cite{Gaubatz,Bergmann,Vitanov2001a,Vitanov2001b,Mackie,Pu} Longhi
proposed \cite{Longhi} a scheme in which the fundamental frequency
field is directly converted into the third harmonic without a
transient generation of the second harmonic. This proposal
requires the simultaneous phase matching of second harmonic
generation ($\omega + \omega =2\omega$) and sum frequency
generation ($\omega +2\omega =3\omega$); because this condition
can be fulfilled only for a specific frequency this technique is
not broadband. Suchowski \emph{et al.}
\cite{Suchowski2008,Suchowski2009} used an aperiodically poled QPM
crystal to achieve both high efficiency and large bandwidth in sum
frequency generation (SFG) in the undepleted pump approximation
using ideas from rapid adiabatic passage in quantum physics
\cite{Allen,Vitanov2001a}.

In this Letter, we make use of the analogy between coherent population transfer in three-state quantum systems
 and the two simultaneous collinear second-order parametric processes
  $\omega_3=\omega_1+\omega_2$ and $\omega_4=\omega_1+\omega_3=2\omega_1+\omega_2$
 to design a potentially highly efficient broadband SFG technique.
To this end, we use linearly chirped QPM gratings \cite{Lefort1,Lefort2,Arbore}, which provide the analogy to level crossings in atomic systems \cite{Broers,Unanyan,Ivanov}.

The two simultaneous SFG processes $\omega_3=\omega_1+\omega_2$ and $\omega_4=\omega_1+\omega_3$, for %an aperiodically poled
 a QPM crystal with susceptibility $\chi^{(2)}$ and local modulation period $\Lambda (z) $ are described by the set of nonlinear differential equations \cite{Saltiel,Arie}
\begin{subequations}\label{nonlinear system}
\begin{align}
i\dz E_1 &= \Omega_1\left( E_2^{\ast }E_3e^{-i\Delta_1z}+E_3^{\ast }E_4e^{-i\Delta_2z}\right) ,  \label{E1} \\
i\dz E_2 &= \Omega_2E_1^{\ast }E_3e^{-i\Delta_1z}, \label{E2} \\
i\dz E_3 &= \Omega_3\left( E_1E_2e^{i\Delta_1z}+E_1^{\ast }E_4e^{-i\Delta_2z}\right) ,  \label{E3} \\
i\dz E_4 &= \Omega_4E_1E_3e^{i\Delta_2z},  \label{E4}
\end{align}
\end{subequations}
where $z$ is the position along the propagation axis, $c$ is the speed of light in vacuum, and $E_j$, $\omega_{j}$ and $n_{j}$ are the electric field, the frequency and the refractive index of the $j$-th laser beam, respectively.
Here $\Omega_{j}=\chi^{\left( 2\right) }\omega_{j}/4cn_{j}$ ($j=1,2,3,4$) are the coupling coefficients,
 while $\Delta_1=n_1/\left( c\omega_1\right) +n_2/\left( c\omega_2\right) -n_3/\left( c\omega_3\right) +2\pi /\Lambda $ and $\Delta_2=n_1/\left( c\omega_1\right) +n_3/\left( c\omega_3\right)-n_4/\left( c\omega_4\right) +2\pi /\Lambda $
 are the phase mismatches for the $\omega_3$ and $\omega_4$ SFG processes.

\textbf{Undepleted pump approximation.}
The coupled nonlinear equations \eqref{nonlinear system} are often linearized assuming that the incident pump field $E_1$ is much stronger than the other fields and therefore its amplitude is nearly constant (undepleted) during the evolution.
Then Eqs.~\eqref{nonlinear system} are reduced to a system of three linear equations,
\begin{equation}
i\dz \textbf{A}(z) = \M(z) \textbf{A}(z),\quad \M =
\left[\begin{array}{ccc}
-\Delta_1 & \Omega_{p}^* & 0 \\
\Omega_{p} & 0 & \Omega_{s}^* \\
0 & \Omega_{s} & \Delta_2
\end{array}\right]
\label{three states system}
\end{equation}
with $\Omega_{p}=E_1\sqrt{\Omega_2\Omega_3}$, $\Omega_{s}=E_1\sqrt{\Omega_3\Omega_4}$, $\mathbf{A}(z) = [A_2(z),A_3(z),A_4(z)]^T$, $A_2=E_1 E_2 e^{i\Delta_1 z}\sqrt{\Omega_3\Omega_4/2}$, $A_3=E_1E_3\sqrt{\Omega_2\Omega_4/2}$, $A_4=E_1 E_4 e^{-i\Delta_2 z}\sqrt{\Omega_2\Omega_3/2}$.
Upon the substitution $z\to t$, Eq.~\eqref{three states system} becomes identical to the time-dependent Schr\"{o}dinger equation
 for a three-state quantum system in the rotating-wave approximation, which is studied in great detail \cite{Vitanov2001a};
 the vector $\mathbf{A}(z)$ and the driving matrix $\M$ correspond to the quantum state vector and the Hamiltonian, respectively.
We note that the quantity $|\mathbf{A}(z)|^2 = |A_2(z)|^2+|A_3(z)|^2+|A_4(z)|^2$ is conserved, like the total population in a coherently driven quantum system.
By definition, in the adiabatic regime the system stays in an eigenvector of the ``Hamiltonian'' $\M$.
%When the local modulation period $\Lambda(z)$ is made to vary linearly along $z$
We assume that $\Delta_1(z)$ and $\Delta_2(z)$ change linearly along $z$, %; $\Delta_1(z)=\delta\pm\alpha^2 z$ and $\Delta_2(z)=-\delta\pm\alpha^2 z$,
 which can be achieved, for example, by varying $\Lambda(z)$.
Explicitly, we assume that either $\Delta_1=\delta - \alpha^2z$ and $\Delta_2=-\delta - \alpha^2z$, which is called ``intuitive sweep" (for reasons that will becomes clear shortly)
 or $\Delta_1=\delta + \alpha^2z$ and $\Delta_2=-\delta + \alpha^2z$ which is called ``counterintuitive sweep''.
For the sake of generality, we take hereafter $\alpha$ as the unit
of coupling and $1/\alpha$ as the unit of length. Then the three
eigenvalues of $\M$ will cross each other at three different
distances $z_m$ ($m=1,2,3$), thereby creating a triangle crossing
pattern \cite{Broers,Unanyan,Ivanov}. These crossings allow us to
design recipes for efficient broadband SFG, in analogy to
adiabatic passage techniques in quantum physics
\cite{Allen,Vitanov2001a,Vitanov2001b,Broers,Unanyan,Ivanov}.
Because of the analogy to the Schr\"odinger equation the condition
for adiabatic evolution can be derived using the
Landau-Zener-Majorana model \cite{Landau,Zener,Majorana} and reads
(for linear chirping and constant couplings): $|\Omega_{x}|
\gtrsim \alpha$, where $\Omega_x$ is the relevant coupling at the
respective crossing.

%***************************************************************
\begin{figure}[t]
\centerline{\includegraphics[width=7.5cm]{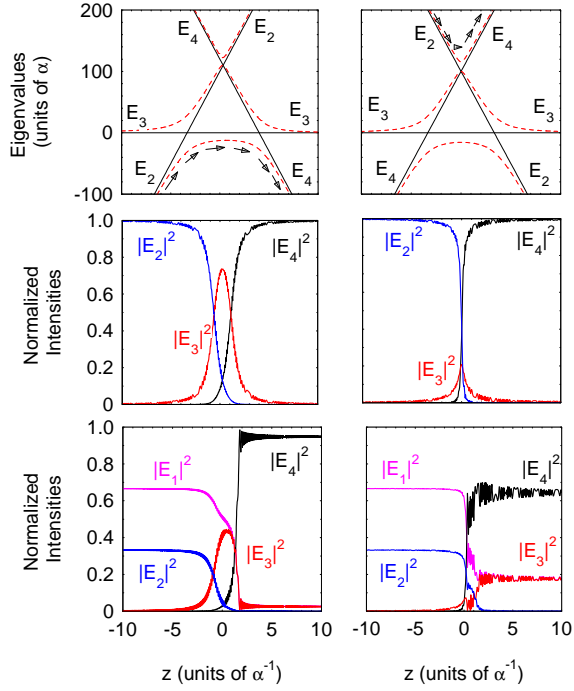}}
\caption{(Color online) Sequential SFG of $\omega_4$ field.
Top frames: Diagonal elements (solid lines) and eigenvalues (dashed lines) of the driving matrix $\M$ of Eq.~\eqref{three states system}
 for the ``intuitive'' (left frames) and ``counterintuitive'' (right frames) phase mismatch sweep.
The field intensities are calculated numerically from Eqs.~\eqref{nonlinear system} for $\delta =2\alpha $, $\Omega_1=\Omega_2=\Omega_3=\Omega_4=\alpha$.
Middle frames: undepleted pump, $|E_1(z_i)|^2 = 100|E_2(z_i)|^2$, with $z_i=-20\alpha^{-1}$;
bottom frames: depleted pump, $|E_1(z_i)|^2 = 2|E_2(z_i)|^2$.
}
\label{Fig1}
\end{figure}
%***************************************************************

Figure \ref{Fig1} plots the eigenvalues of $\M$ of Eq.~(\ref{three states system}) vs $z$. % for constant $\Omega_{p}$ and $\Omega_{s}$.
Initially only the $\omega_2$ field is present, hence the vector $\mathbf{A}=[A_2,0,0]$.
If the evolution is adiabatic then there are two possible paths that the system can follow (marked by arrows).
If the phase match for the $\omega_3$ generation process occurs first (left frames of Fig.~\ref{Fig1}), then the energy is converted first to the $\omega_3$ field and then to the $\omega_4$ field.
This ``intuitive'' two-step scheme extends the single-step adiabatic passage scenario for SFG \cite{Suchowski2008,Suchowski2009}.
Interestingly, we find that efficient energy transfer directly to the $\omega_4$ field is also possible through the ``counterintuitive'' direction
 of the local modulation period sweep when the phase match for the $\omega_4$ generation process occurs first (right frames of Fig.~\ref{Fig1}).
Then the energy flows from the $\omega_2$ field to the $\omega_4$ field with almost no energy transferred to the intermediate $\omega_3$ field.

%***************************************************************
\begin{figure}[t]
\centerline{\includegraphics[width=7.5cm]{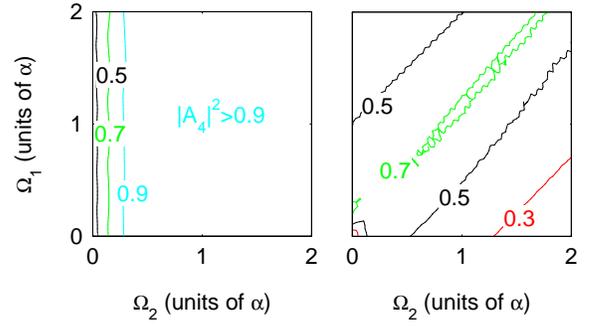}}
\caption{(Color online)
Efficiency of SFG of $\omega_4$ field vs the couplings $\Omega_1$ and $\Omega_2$ obtained by numerical integration of Eqs.~\eqref{nonlinear system}
 for ``counterintuitive sweep'' with $\delta=\alpha$ and $\Omega_3=\Omega_4=\alpha$.
Left frame: undepleted pump, $|E_1(z_i)|^2 = 100|E_2(z_i)|^2$, with $z_i=-20\alpha^{-1}$;
right frame: depleted pump, $|E_1(z_i)|^2 = 2|E_2(z_i)|^2$.
}
\label{Fig2}
\end{figure}
%***************************************************************

\textbf{Depleted pump.}
We have found by numerical integration of the nonlinear system \eqref{nonlinear system} that the described scheme is also applicable beyond the undepleted pump approximation, when the $\omega_1$ and $\omega_2$ fields have comparable energies; this is demonstrated in the bottom frames of Fig.~\ref{Fig1}.
Unfortunately, many optical parametric processes such as $|\omega_1-\omega_2|$, $2\omega_1$, $2\omega_2$, $\omega_1+2\omega_2$ become possible in this case and it is not easy to find the conditions for broadband SFG of the $\omega_4$ field.

The contour plot in Fig.~\ref{Fig2} demonstrates the robustness of SFG of the $\omega_4$ field against parameter variations.
SFG for an undepleted pump (left frame) is remarkably robust in confirmation of the simple analytic theory described above.
SFG for a depleted pump (right) is less robust although relatively high SFG efficiency is still possible;
 because then the simple eigenvalue arguments cannot be used the interpretation is more difficult.

\textbf{Third harmonic generation.}
Third harmonic generation is an important special case of SFG, which is readily treated in the adiabatic regime.
The respective equations are derived from Eqs.~\eqref{nonlinear system},
\begin{subequations}\label{THG}
\begin{align}
i\dz A_{\omega } &= \Omega_{\omega }A_{\omega }^{\ast }A_{2\omega}e^{-i\Delta_1z}+\Omega_{2\omega }A_{2\omega }^{\ast }A_{3\omega}e^{-i\Delta_2z}, \\
i\dz A_{2\omega } &= \Omega_{\omega }A_{\omega }^2e^{i\Delta_1z}+\Omega_{2\omega }A_{\omega }^{\ast }A_{3\omega }e^{-i\Delta_2z},\\
i\dz A_{3\omega } &= \Omega_{2\omega }A_{\omega }A_{2\omega}e^{i\Delta_2z},
\end{align}
\end{subequations}
where $\Omega_{\omega}={\chi^{(2)}\omega}/{4c n_{\omega}}$,
$\Omega_{2\omega}={\chi^{(2)}\omega
\sqrt3}/{4c\sqrt{n_{2\omega}n_{3\omega}}}$,
 $E_1=E_2$, $A_{\omega}=E_1\sqrt{{2n_{\omega }}/{n_{2\omega }}}$, $A_{2\omega}=E_3$, $A_{3\omega}=E_4\sqrt{{2n_{3\omega}}/{3n_{2\omega }}}$.

%***************************************************************
\begin{figure}[t]
\centerline{\includegraphics[width=7.5cm]{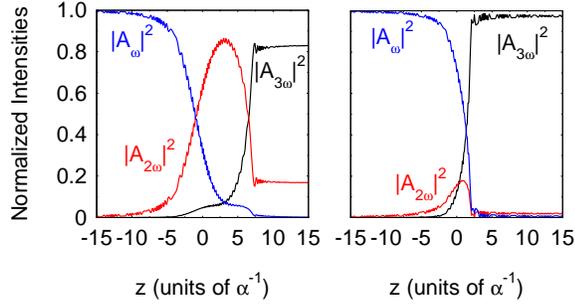}}
\caption{(Color online) Third harmonic generation for ``intuitive'' (left frames)
 and ``counterintuitive'' (right frames) phase mismatch sweep calculated numerically from Eqs.~\eqref{nonlinear system}
 for $\delta =4\alpha $, $\Omega_{\omega}=2\alpha$, $\Omega_{2\omega}=5\alpha$.
}
\label{Fig3}
\end{figure}
%***************************************************************

\begin{figure}[t]
\centerline{\includegraphics[width=7.5cm]{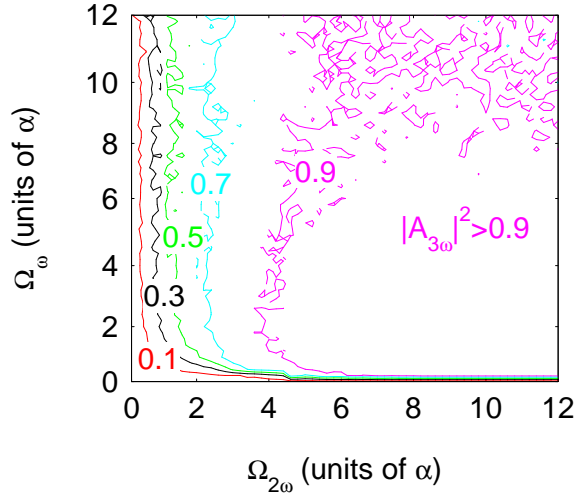}}
\caption{(Color online) Efficiency of third harmonic generation efficiency vs the couplings $\Omega_{\omega}$ and $\Omega_{2\omega}$ for ``counterintuitive'' sweep
 obtained by numerical integration of Eqs.~\eqref{nonlinear system} for $\delta=2\alpha$.
 }
\label{Fig4}
\end{figure}

Figure \ref{Fig3} shows numerical simulation of third harmonic generation.
There are again two possible scenarios.
If the phase match for the second harmonic generation process occurs first (``intuitive'' sweep, left frame of Fig.~\ref{Fig3}),
 then the efficiency of the third harmonic is good, but we have some unwanted second harmonic left.
For the ``counterintuitive'' direction of the local modulation period sweep, when the phase match for the third harmonic generation occurs first,
 the second harmonic is strongly suppressed and a nearly complete transfer of energy to the third harmonic takes place (right frame of Fig.~\ref{Fig3}).
Figure \ref{Fig4} demonstrates the robustness of the third harmonic generation for the ``counterintuitive'' phase mismatch sweep,
 which indicates that this technique is broadband.

%***************************************************************
%***************************************************************

\textbf{Conclusion.}
We have used the analogy between the time-dependent Schr\"odinger equation and the SFG equations in the undepleted pump approximation to propose an efficient broadband SFG technique.
A local modulation period sweep along the light propagation creates crossings in the phase matching between different parametric processes,
 which in combination with adiabatic evolution conditions allow efficient and robust SFG of the desired frequency $\omega_4=2\omega_1+\omega_2$.
While the physical picture is transparent in the undepleted pump approximation, the basic feature of the SFG process remain largely intact in the general regime of depleted pump.
Chirped QPM gratings offer robustness against variations of the parameters of both the crystal and the electric fields,
 which include the crystal temperature, the wavelengths of the input electric fields, the crystal length and the angle of incidence.

The present work can  be viewed as a generalization of the idea of Suchowski \emph{et al.} \cite{Suchowski2008} from a single SFG to simultaneous SFG processes in and beyond the undepleted pump approximation.
This work is also a broadband alternative to the (narrowband) STIRAP-based third harmonic generation proposal of Longhi \cite{Longhi}.

This work is supported by the European network FASTQUAST and the Bulgarian NSF grants D002-90/08 and DMU02-19/09. %and Sofia University Grant 022/2011.

%%%%%%%%%%%%%%%%%%%%%%%%%%%%%%%%%%%%%%%%%%%%%%%%%%%%%%%%%%%%%%%%%%%%%%%%%%%%%%%%%%%%%%%%%%%%%%%%%%%%%%%%%%%%%%%%%%%%%%%%%%%%%%%%%%%%%%%%%%%%%%%%%%%%


\begin{thebibliography}{99}
\bibitem{Saltiel} S. M. Saltiel, A. A. Sukhorukov, and Y. S. Kivshar,
\textquotedblleft Multistep parametric processes in nonlinear
optics\textquotedblright , Prog. Opt. \textbf{47}, 1-73 (2005).

\bibitem{Arie} A. Arie and N. Voloch, \textquotedblleft Periodic,
quasi-periodic, and random quadratic nonlinear photonic
crystals\textquotedblright , Laser and Photon. Rev. \textbf{4},
355-373 (2010).

\bibitem{Longhi} S. Longhi, \textquotedblleft Third-harmonic generation in
quasi-phase-matched $\varkappa ^{\left( 2\right) }$ media with
missing second harmonic\textquotedblright , Opt. Lett.
\textbf{32}, 1791-1793 (2007).

\bibitem{Suchowski2008} H. Suchowski, D. Oron, A. Arie, Y. Silberberg,
\textquotedblleft Geometrical representation of sum frequency generation and adiabatic frequency conversion\textquotedblright , Phys. Rev. A. \textbf{78}%
, 063821 (2008).

\bibitem{Suchowski2009} H. Suchowski, V. Prabhudesai, D. Oron, A. Arie, Y.
Silberberg, \textquotedblleft Robust adiabatic sum frequency
conversion\textquotedblright , Opt. Express \textbf{17},
12731-12740 (2009).

\bibitem{Gaubatz} U. Gaubatz, P. Rudecki, S. Schiemann, K. Bergmann,
\textquotedblleft Population transfer between molecular
vibrational levels by stimulated Raman scattering with partially
overlapping laser fields. A new concept and experimental
results\textquotedblright , J. Chem. Phys. \textbf{92}, 5363-5377
(1990).

\bibitem{Bergmann} K. Bergmann, H. Theuer, B. W. Shore, \textquotedblleft
Coherent population transfer among quantum states of atoms and
molecules\textquotedblright , Rev. Mod. Phys. \textbf{70},
1003-1025 (1998).

\bibitem{Vitanov2001a} N. V. Vitanov, T. Halfmann, B. W. Shore, and K.
Bergmann, \textquotedblleft Laser-induced population transfer by
adiabatic passage techniques\textquotedblright , Annu. Rev. Phys.
Chem. \textbf{52}, 763-809 (2001).

\bibitem{Vitanov2001b} N. V. Vitanov, M. Fleischhauer, B. W. Shore, and K.
Bergmann, \textquotedblleft Coherent manipulation of atoms and
molecules by
sequential laser pulses\textquotedblright , Adv. At. Mol. Opt. Phys. \textbf{%
46}, 55-190 (2001).


\bibitem{Mackie} M. Mackie, R. Kowalski, and J. Javanainen,
\textquotedblleft Bose-stimulated Raman adiabatic passage in
photoassociation\textquotedblright , Phys. Rev. Lett. \textbf{84},
3803-3806 (2000).

\bibitem{Pu} H. Pu, P. Maenner, W. Zhang, and H. Y. Ling, \textquotedblleft
Adiabatic condition for nonlinear systems\textquotedblright ,
Phys. Rev. Lett. \textbf{98}, 050406 (2007).

\bibitem{Allen} L. Allen, J. H. Eberly, \emph{Optical Resonance and
Two-Level Atoms} (Dover, New York, 1987).

\bibitem{Lefort1} M. Charbonneau-Lefort, B. Afeyan, and M. M. Fejer,
\textquotedblleft Optical parametric amplifiers using chirped
quasi-phase-matching gratings I: practical design
formulas\textquotedblright , J. Opt. Soc. Am. B \textbf{25},
463-480 (2008).

\bibitem{Lefort2} M. Charbonneau-Lefort, M. M. Fejer, and B. Afeyan,
\textquotedblleft Tandem chirped quasi-phase-matching grating
optical parametric amplifier design for simultaneous group delay
and gain control\textquotedblright , Opt. Lett. \textbf{30},
634-636 (2005).

\bibitem{Arbore} M. A. Arbore, O. Marco, and M. M. Fejer, \textquotedblleft
Pulse compression during second-harmonic generation in aperiodic
quasi-phase-matching gratings\textquotedblright , Opt. Lett.
\textbf{22}, 865-867 (1997).

\bibitem{Broers}  B. Broers, L. D. Noordam, and H. B. van Linden van den
Heuvell, \textquotedblleft Diffraction and focusing of spectral
energy in multiphoton processes\textquotedblright , Phys. Rev. A
\textbf{46}, 2749-2756 (1992).

\bibitem{Unanyan} R. G. Unanyan, N.V. Vitanov, and K. Bergmann,
\textquotedblleft Preparation of entangled states by adiabatic
passage\textquotedblright , Phys. Rev. Lett. \textbf{87}, 137902
(2001).

\bibitem{Ivanov} S. S. Ivanov and N. V. Vitanov, \textquotedblleft Steering
quantum transitions between three crossing energy
levels\textquotedblright , Phys. Rev. A \textbf{77}, 023406
(2008).

\bibitem{Landau} L. D. Landau, \textquotedblleft On the theory of transfer of energy at
collisions II \textquotedblright , Physik Z. Sowjetunion
\textbf{2}, 46-52 (1932).


\bibitem{Zener} C. Zener, \textquotedblleft Non-adiabatic crossing of energy levels
\textquotedblright , Proc. R. Soc. Lond. Ser. A \textbf{137},
696-702 (1932);

%\bibitem{St\"{u}ckelberg} E. C. G. St\"{u}ckelberg, Helv. Phys. Acta \textbf{5}, 369 (1932).

\bibitem{Majorana} E. Majorana, \textquotedblleft Atomi orientati in campo magnetico variabile \textquotedblright , Nuovo Cimento \textbf{9}, 43-50 (1932).

\end{thebibliography}

\newpage
\end{document}